\documentclass[conference]{IEEEtran}
\IEEEoverridecommandlockouts
\usepackage{cite}
\usepackage{amsmath,amssymb,amsfonts}
\usepackage{algorithmic}
\usepackage{graphicx}
\usepackage{soul}
\usepackage{subfig}
\usepackage{textcomp}
\usepackage{xcolor}
\def\BibTeX{{\rm B\kern-.05em{\sc i\kern-.025em b}\kern-.08em
    T\kern-.1667em\lower.7ex\hbox{E}\kern-.125emX}}

\begin{document}

\title{Deploying AI-Based Applications with Serverless Computing in 6G Networks: An Experimental Study\\
}

\author{\IEEEauthorblockN{~Marc~Michalke, ~Chukwuemeka~Muonagor ~and~Admela~Jukan}
    \IEEEauthorblockA{\\ \textit{Institute of Computer and Network Engineering,}
        \textit{Technische Universit{\"a}t Braunschweig}, Germany. \\
        E-mail: \{m.michalke, c.muonagor, a.jukan\}@tu-bs.de}
}

\maketitle

\begin{abstract}
Future 6G networks are expected to heavily utilize machine learning capabilities in a wide variety
of applications with features and benefits for both, the end user and the provider. While the
options for utilizing these technologies are almost endless, from the perspective of network
architecture and standardized service, the deployment decisions on where to execute the AI-tasks are
critical, especially when considering the dynamic and heterogeneous nature of processing and
connectivity capability of 6G networks. On the other hand, conceptual and standardization work is
still in its infancy, as to how to categorizes ML applications in 6G landscapes; some of them are
part of network management functions, some target the inference itself, while many others emphasize
model training. It is likely that future mobile services may all be in the AI domain, or combined
with AI. This work makes a case for the serverless computing paradigm to be used to this end. We
first provide an overview of different machine learning applications that are expected to be
relevant in 6G networks. We then create a set of general requirements for software engineering
solutions executing these workloads from them and propose and implement a high-level edge-focused
architecture to execute such tasks. We then map the ML-serverless paradigm to the case study of 6G
architecture and test the resulting performance experimentally for a machine learning application
against a setup created in a more traditional, cloud-based manner. Our results show that, while
there is a trade-off in predictability of the response times and the accuracy, the achieved median
accuracy in a 6G setup remains the same, while the median response time decreases by around 25\%
compared to the cloud setup.
\end{abstract}


\section{Introduction}

With the development of 6G in full progress, visions on what this new technology will bring are constantly shifting and changing. Some projections however already reached a wide consensus; particularly the use of artificial intelligence (AI) and machine learning (ML) in the setup, maintenance and management will be a key component of 6G systems \cite{9493388}. With help of AI and ML-enabled intelligence, 6G will be able to support enhanced services, such as Mobile Broadband communications (eMBB), Ultra-reliable Low Latency Communications (URLLC) and massive Machine Type Communications (mMTC). ML techniques can also be utilized in various wireless communication problems, which include optimization, classification, analysis and signal recovery \cite{10396677}. It is also predicted that 6G is  expected to shift to a more flat and distributed hierarchy~\cite{baldoni_data-centric_2023}, as well as relying on more softwareized and cloudified infrastructure. Many cloud technologies and tools, - now fully emerging in the edge computing, are now being adopted for mobile computing.  One of these technologies is the serverless computing paradigm,  i.e., a workload distribution among multiple nodes, with latest research also showing that it can be used effectively at the mobile edge and constrained 6G resources\cite{sabbioni_serverless_2024}. This joint evolution of networking and cloud computing towards serverless computing paradigm is significant, as it can potentially benefit both the network architectures and the  target computing applications in mobile networks \cite{serrano_guest_2024}. 

We are especially interested in the ability of serverless computing paradigm in 6G networks to provision AI based services and increase machine learning efficiency, as previous work already recognized in the cloud realm~\cite{sanchez-artigas_experience_2021}. In the context of 6G, current research is primarily focused on how ML can be applied to 6G landscapes on a functionality level, or targets the inference itself, while others emphasize on model training. However, the fact that the 6G communication systems are expected to operate with extremely low latency \cite{10209158} is a challenge that training of ML models in the 6G architecture faces due to the delay caused by the training process.
The execution of the training therefore remains an open issue, since traditional approaches, like transferring all the training data to the cloud, would result in a massive communication overhead as well as potential privacy concerns, depending on the nature of the data.
An additional problem faced in this emerging network is resource utilization. With access network devices being expensive to replace or upgrade, the providers try to limit CAPEX where possible. An appropriate utilization of the available resources is therefore crucial.

In this work, we study experimentally the features of the serverless computing paradigm to
accommodate the classes of ML applications emerging in 6G networks, and map the ML/serverless
architecture to a typical 6G scenario. To this end, we first collect potential applications for
machine learning in 6G and define a set of objectives that a software stack should meet. We then
propose a generic architecture that can be implemented on top of a broad range of devices, and show
how typical ML applications can be mapped to serverless mobile computing. Finally, we experimentally
study on how it could benefit a real-world scenario and test its performance for an exemplary
machine learning application in a typical 6G scenario against a setup created in a more traditional,
cloud-based manner. Our results show that, while there is a trade-off in predictability of the
response times, and accuracy, the achieved median accuracy of the application's predictions remains
the same in a 6G setup, while the median response time decreases by around 25\% compared to the
cloud setup. Finally, to highlight the potential advantages and disadvantages of employing the
serverless paradigm to support machine learning applications in upcoming 6G infrastructures, we make
recommendations for future research.

The remainder of the paper is structured as follows; Section II reviews classes of emerging AI/ML
applications in 6G, Section III proposes a mapping of ML applications to serverless computing.
Section IV furthermore maps serverless computing to a 6G architecture. Section V describes our
experimental study and its results. Section VI concludes the paper.

\section{Classes of AI/ML Applications in 6G}\label{sec:ml-in-6g}

Various ML techniques have already been proposed as potential solutions to upcoming 6G wireless
communication challenges e.g., resource allocation optimization, enhanced network security and
privacy, or latency requirements, with many of them being strong contenders for the application of
federated or distributed learning, while the traditional approach of centralized learning is mostly
considered unsuitable~\cite{muscinelli_overview_2022}. In this section, these 6G requirements and
few examples of ML applications proposed to handle them are reviewed. It should be noted that these
applications are by no means exhaustive, they only serve as motivation to derive features suitable
for serverless computing, as proposed in the sections that follow.

\subsection{Energy Efficiency and Capacity Optimisation}
Intelligent and self-healing machines, including autonomous vehicles and connected drones, require
communication links in Device-to-Device (D2D) communication and connectivity with base stations,
which increases the concern of energy consumption about sustainable wireless infrastructure
\cite{10209158}. Here, an approach that used supervised machine learning was studied for cache
localization in D2D communications, which reduces the access delay of User Equipment (UE) and
minimizes energy consumption while predicting accurate cache placement locations. Energy efficiency has furthermore significance in designing 6G devices due to their utilization of higher frequency bands, that comparably consume more energy.
 
\subsection{Mobility and Handover Management}
Handover management has a prime significance in mobile networks for maintaining QoS due to several
challenges like throughput reduction and service disruption \cite{10209158}. Deep Learning (DL) can
be used to enhance mobility management performance in 6G networks by predicting user mobility
patterns, channel stability and signal strength, leading to better handover performance and reduced
network congestion. Paper  \cite{10200390} proposes a DNN model that accurately predicts the next
base station, enhancing the robustness of the conditional handover based on the signal patterns of
the base station and the received power of the reference signal.

\subsection{Resource Allocation Optimisation}\label{sec:res-opt}
Intelligent resource management is a key feature of 6G networks that enables self-configuration and
self-healing and enhances energy efficiency by leveraging ML techniques for parallel computing and
autonomous decision-making \cite{10209158}.  Paper \cite{10489587} proposes Machine Learning-Based
Resource Allocation Algorithms for 6G Networks. They examined series of ML techniques ranging from
regression, SVMs to DNN to optimise and manage radio resources in cellular networks based on data
like interference constraints, user preferences, and network performance. Their solutions
appropriately balances network efficiency and user QoS. Another metric they relied on here is
network traffic classification, which has already been discussed in-depth for traditional networks
\cite{pacheco_towards_2019}.

\subsection{Latency and Reliability Requirements}
The upcoming 6G communication systems are designed to operate at terabit-per-second data rates with
extremely low latency. Massive MTCs and URLLCs have revolutionized cellular communications recently
\cite{10209158}. Paper \cite{10105173} discusses how to implement ML algorithms to guarantee the QoS
requirements for different URLLC scenarios, including mobility URLLC, massive URLLC, and broadband
URLLC. They also presented a case study of downlink URLLC channel access problems, solved by
centralized deep reinforcement learning (CDRL) and federated DRL (FDRL), respectively, validating
the effectiveness of ML for URLLC services.

\subsection{Data Privacy}
Also data privacy can be improved with AI/ML. Paper \cite{liu2020federated} proposed federated
learning (FL) in 6G to handle the requirements of centralized data collection and processing by a
central server which is a bottleneck of large-scale implementation due to significantly increasing
privacy concerns.

\subsection{Network Security}\label{sec:net-sec}
AI/ML technologies have been proposed to detect different kinds of attacks in the 6G architecture.
For instance, \cite{10246078} proposed a method for Attacks Detection in 6G (AD6Gs) wireless
networks using Random-Forest and Support Vector Machine algorithms for classification based on the
traffic and communication pattern. The proposed method outperformed the existing classification
methods. Paper \cite{10333795} presented a novel deep learning-based approach to detect various
cyber attacks within 6G wireless networks, encompassing DoS, probe attacks, and Sybil attacks.
Leveraging the KDD Cup dataset and implementing their solution using PyTorch, the method
demonstrates remarkable effectiveness, surpassing conventional techniques and adaptability to
evolving attack patterns.

\subsection{Task Offloading}
A significant number of latency-aware and computation-intensive tasks will have to be executed in 6G
wireless networks due to multiple radio access technologies, slices, servers, and upcoming vehicular
applications. The ability to make real-time offloading decisions is a major challenge in reducing
latency and energy consumption \cite{10209158}. In \cite{wu2020collaborate}, a Distributed Deep
learning-driven Task Offloading (DDTO) algorithm was presented to make offloading decisions in
heterogeneous networks. It adaptively modifies parameters to make near-optimal offloading decisions
by learning from past offloading events in MEC and Mobile Cloud Computing (MCC) heterogeneous
environments.

\section{Mapping ML Applications to Serverless Computing Paradigm}

In this section, we propose a method to mapping the ML applications to a sample serverless
architecture, which can be used for distributed learning in mobile networks. We only focus on the
training aspects here, which, by nature, take significant time and, while it can usually not happen
in real time, should be reduced where possible. The inference however, is particularly
time-sensitive and will likely happen locally on devices, supported by neural processing units
(NPUs) to avoid the communication overhead that offloading would introduce. This on the other hand
would require a separate study. Our goal here is to show the engineering process of ML mapping to
serverless paradigm, while opening avenues for a more broadened scope in future networks.

\subsection{Software Engineering Objectives}
For our architecture, we define the following objectives to be met for the approach to be feasible
for as many of the previously explained expected uses of ML in 6G as possible:
\begin{itemize}
    \item Compatibility: allow for easy integration with state-of-the-art edge computing solutions
    \item Scalability: allow for quick horizontal and vertical scaling through stateless operation
    \item Stability: utilize only mature and tested tools
    \item Versatility: allow for easy adaptation to a variety of ML models, problems and solutions
\end{itemize}

\subsection{Architecture and Implementation}
For any distributed ML application to be deployable on top of a serverless, or a Function-as-a-Service
(FaaS) platform, it first has to be transformed into a serverless function. In this section, we will
start by briefly introducing an example ML application for distributed learning which trains an ML
model with text examples for classification, which is a task almost identical in nature to the network traffic
classification application presented in Sections \ref{sec:net-sec} and \ref{sec:res-opt}. We then
follow up by preparing the code for encapsulation into a serverless function and the actual
containerization process through \emph{Docker}. The FaaS framework used (here: \emph{Knative}) is
already integrated into an open-source testbed called \emph{Benchfaster}, which we use here to
create and implement a serverless 6G network testbed. It should be noted that the \emph{Benchfaster}
code, freely and openly available for use and experimentation \cite{Carpio2022}, enables automated deployment and benchmarking of containerized applications over real-world networking hardware, able to provision the devices with all needed software prerequisites. The full code of our ML application is available online at Github\footnote{F. Carpio "Serverless Functions" GitHub, 2024. [Online]. Available: https://github.com/fcarp10/serverless-functions}.

\begin{figure}[htbp]
    \centerline{\includegraphics[width=0.4\textwidth]{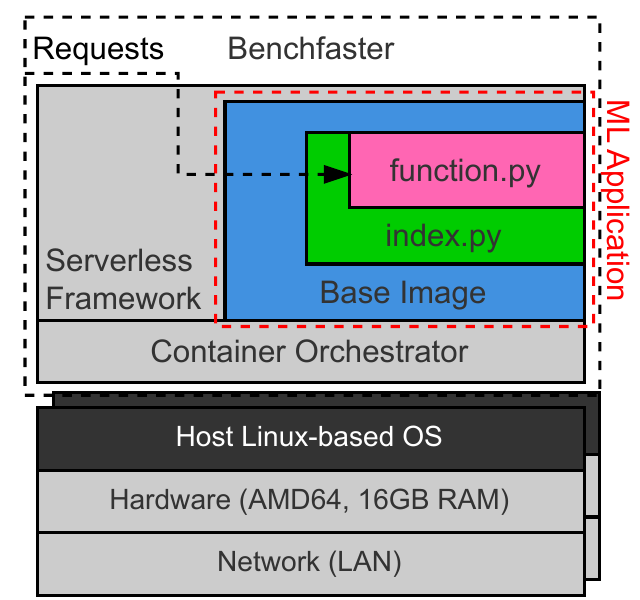}}
    \caption{Example Architecture}
    \label{fig:arch}
\end{figure}

We adapt the Benchfaster testbed architecture to map ML functions to the serverless paradigm and furthermore to test the sample 6G use cases. First, we propose the architecture depicted in Fig.~\ref{fig:arch}. It consists of several main components which will be explained in the remainder of this section.

\subsubsection{Container Orchestrator}
To execute tasks like container lifecycle and requirement management, a lightweight Kubernetes
distribution for container orchestration should be used, partially because of Kubernetes' wide
adoption for cloud use cases, making it a very mature and well supported solution and partially
because several light-weight distributions of it are currently emerging, all targeting devices with
lower compute capacity than the cloud; edge devices for example.

In our example, we chose \texttt{k3s} for this task since we have preexisting experience with it,
there is no strong argument currently to be made for this solution over any of the competitors like
\texttt{microk8s}, \texttt{k0s} or similar. The figure hints that this architecture can be deployed
across multiple devices, which can be done through the creation of a Kubernetes cluster, allowing
for more flexible workload distribution.

\subsubsection{Serverless Framework}
A serverless framework is used to provide the necessary components to run serverless functions,
e.g., an HTTP endpoint. Incoming requests are distributed among the available replicas, each of
which then executes the training algorithm and returns the result to the requesting device. Here, we
use \texttt{Knative}, the currently most widely used serverless platform, as we deem it the most
representative solution.

\subsubsection{ML Application}
The containerized serverless application consists of a container/pod created from a container image.
Replication of this entity is possible, producing several, identical, stateless so-called replicas
of the function. The container image is created from a base image (Python Debian Slim), providing
needed libraries for the application code, and then complemented with the actual application code.
This code can be written in several programming languages depending on the individual task, which
also influences the choice of the base image.

\subsubsection{function.py}
The actual training implementation is heavily based on the individual use case and can be
implemented in almost any programming language, as long as this is supported by the serverless
framework.

Our function is implemented in Python, since the language provides excellent support for a wide
variety of ML libraries, making it suitable to represent the foundation for a wide variety of
potential use cases. The actual ML code resides in a single file; \texttt{function.py} in this
case, which can vary based on style, the chosen language, and complexity of the function.

\subsection{The Function}\label{application}
The function used  leverages the passive Aggressive Classifier (PAC) model for text classification.
Once a request is received for processing, the \texttt{index.py} calls the function code inside the
file \texttt{function.py}. The input for the machine learning application consists of a json file
with the concrete data consisting of a set of 3150 news texts in combination with their labels which
can be either FAKE or REAL to recognize fake news, due to lack of network-related data. The task of
classification however remains identical with the traffic classification explained in section
\ref{sec:net-sec}, allowing us to apply the results to a real-world 6G use case. This data was
extracted from the news dataset for fake news detection hosted by \emph{Kaggle} on their website
\cite{newscsv}.

To train the models, we chose the method of cross validation, to avoid overfitting
\cite{scikit-learn}. The received data is used to train the model for the number of times specified
by the number of folds for the cross validation. For each fold, a subset of the data is selected as
validation subset, and training is done on the remaining data. By evaluating against the validation
data, the accuracy is then calculated and stored. This process is repeated for the subsequent folds
while ensuring that validation subsets differ every time, resulting in an accuracy value for each
fold. The model parameters leading to the best performance (highest validation accuracy) are
then considered the best parameters and could later be used to train the model on the entire training
dataset. Further details on the PAC model can be found directly at\cite{scikit-learn}.

\section{Mapping Serverless Computing to 6G}
This section will now show how the ML applications running over the serverless computing can be mapped to an example 6G architecture. 
\subsection{6G Architecture}
\begin{figure*}[htbp]
    \centerline{\includegraphics[width=0.85\textwidth]{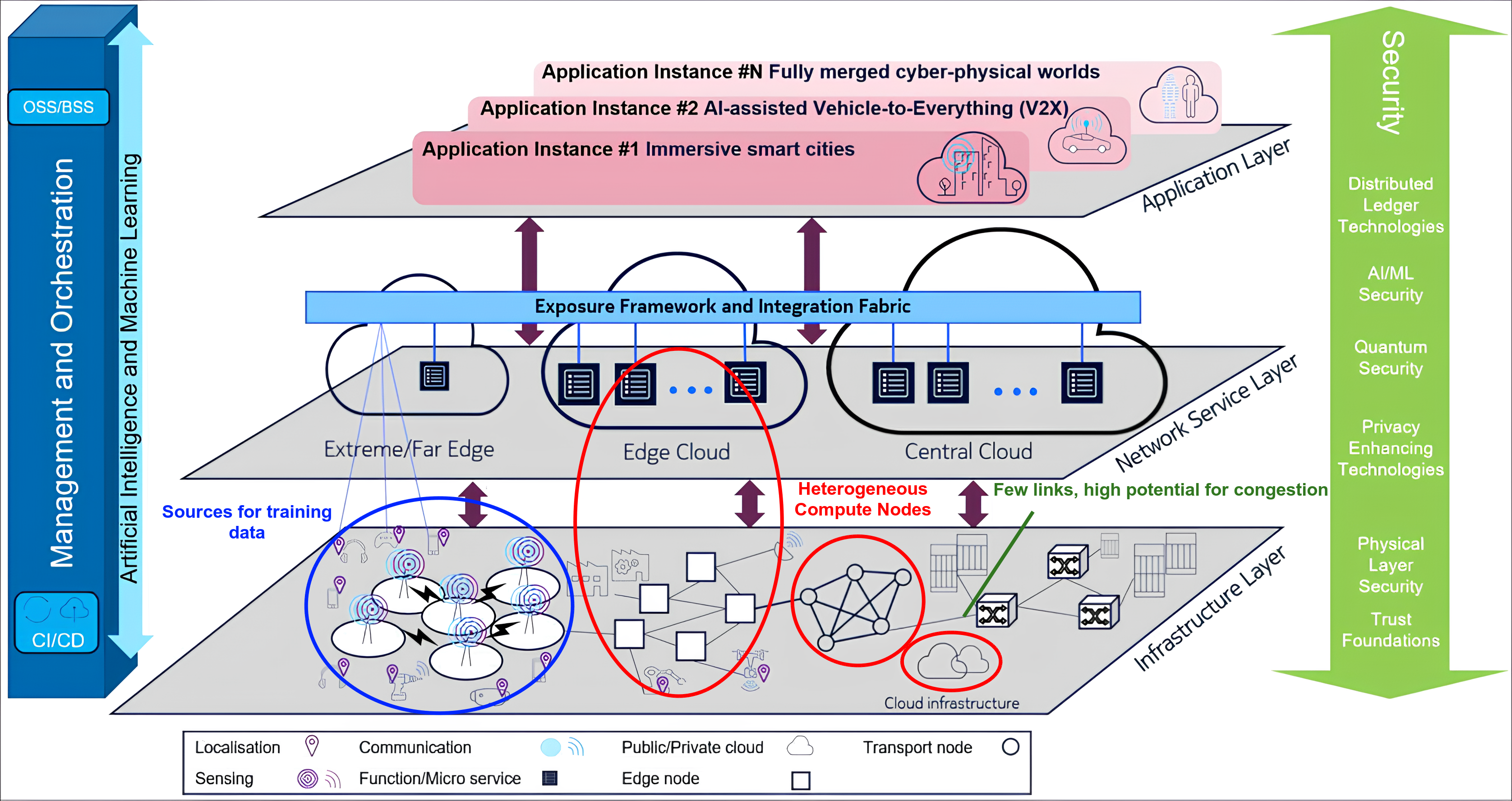}}
    \caption{A 6G architecture as presented by 5G-PPP\cite{bahare_6g_2023}}
    \label{fig:6g-arch}
\end{figure*}

Mapping serverless computing to 6G network is a rather generic task, which we start by first
understanding which devices exist in the 6G landscape to execute ML tasks. To this end, we use one
of the proposed architectures as shown in Fig. \ref{fig:6g-arch} \cite{bahare_6g_2023}. As can be
seen, edge, cloud and transport nodes are envisioned as essential, with the cloud being either
public or private.  Furthermore, transport nodes are connected through a multitude of mesh-like
connections, offering a comparably high link density. It can be assumed that such links would by
nature have less capacity, relative to the experienced traffic, than what we can observe in today's
topologies, which makes it crucial to distribute the load among multiple of them whenever possible.
Similarly, for the processing infrastructure itself, a higher spread of resources typically implies
that the devices have lower individual processing capability, with concrete numbers depending
heavily on the level of technological progress and innovation achieved in the upcoming years and
impossible to anticipate yet. These architectural aspects are central to our idea of deploying ML
applications over serveless computing in 6G networks.

\subsection{The Mapping}
\begin{figure}[htbp]
    \centerline{\includegraphics[width=0.35\textwidth]{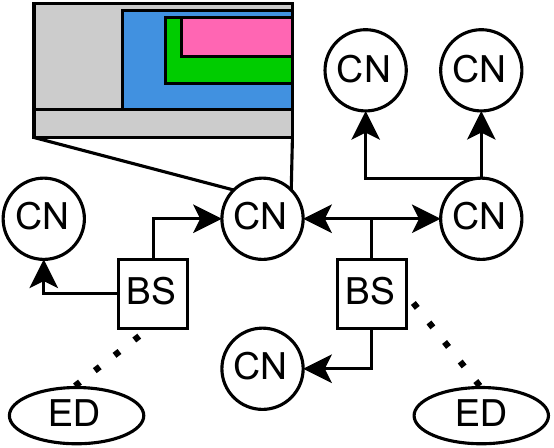}}
    \caption{Mapping to 6G networks}
    \label{fig:6g-mapping}
\end{figure}

To provide a mapping of our proposed architecture to a 6G landscape, we first have to narrow down
the ML application and data that we want to leverage distributed learning for. As explained in
section \ref{sec:ml-in-6g}, a frequently discussed ML use case in 6G is its utilization for traffic
classification for resource optimization or threat/anomaly detection. This means that models have to
be retrained frequently on traffic data and traffic metadata, which have to be collected and
transmitted to the respective training nodes. Resulting in a constant data stream to the cloud in
traditional architectures, leveraging 6G infrastructure nodes and edge computing could mitigate this
problem effectively, as it would not just reduce distance and number of impacted links per
transmission, but would also allow data streams to be spread across multiple horizontal links and
nodes where possible. Lower distances furthermore result in higher data locality, potentially
solving some privacy concerns if the devices are directly controlled by the network operator, as
opposed to some cloud solutions. Given this analysis, Fig. \ref{fig:6g-mapping} proposes an approach
to mapping of the proposed architecture to a 6G landscape.

As it can be seen, we assume a significant number of compute nodes, i.e., edge, cloud and
potentially transport nodes, to be distributed in the network, resulting in multiple training points
for each device that the overall data and load can be distributed upon. This effectively represents
a distributed environment, similar in behavior to topologies already used in other composite
infrastructure like multi-site cloud installations or distributed edge environments.  As
depicted by the arrows, the base stations (BS) would send chunks of their training data, which
they acquired from the wireless connections with the end devices (ED) indicated by the dotted
lines, to the different compute nodes (CN). The compute nodes could either be any hardware
already present in the network for other purposes like running virtual network functions (e.g.
virtual base station), potentially enhanced by accelerator and graphics cards to support ML
tasks, or specifically placed, additional nodes. Specialized nodes could then also be optimized
for the tasks at hand, allowing for higher power efficiency than general purpose devices. 

It would also be possible to leverage devices not directly belonging to the 6G infrastructure, like
external edge nodes or small data centers provided by other application providers, depending on
business-level agreements. Compute nodes also do not all have to be directly reachable by the
offloading device.  The base station in the right-hand side in Fig. \ref{fig:6g-mapping}, could for
example send data to a CN that forms a Kubernetes cluster with two other nodes and forwards requests
to them, allowing for nodes to flexibly join and leave the cluster, especially beneficial for mobile
edge computing where the nodes are changing position and connections. Regardless of what compute
node and hierarchy would be chosen, our proposed software stack can be deployed on top, as indicated
in the figure, as long as the hardware CPU architecture is compatible, which is currently mostly the
case. We can also achieve easy integration with pre-existing compute nodes, since these likely
already employ some sort of container orchestration software, which would be especially beneficial.

It should be noted that the mapping described above focused on the specific ML application of
traffic classification but can be easily applied and repurposed for other 6G ML functionalities as
well as extended at runtime through deployment of different ML functions.

\subsection{Limitations}
One of main feasibility challenges to overcome for our proposed architecture is the suboptimal GPU
support ~\cite{gu_elasticflow_2023}. While it is possible to pass GPU devices and
accelerator cards through to containers via simple volume mounts, this approach does not allow for
sharing, leaving only a single container or pod to be able to use the device exclusively. On some
platforms, GPU sharing can already be achieved, e.g., on Google's GCE or on local machines utilizing
GPU-specific plugins, custom annotations and
CRDs\footnote{https://kubernetes.io/docs/tasks/manage-gpus/scheduling-gpus/}. CPU and memory
resources however can be shared and restricted out of the box in Kubernetes among multiple
containers, allowing for distribution of the resources among all instances. This functionality is
also subject of research for serverless platforms, as can be seen in \cite{satzke_efficient_2020},
and crucial to allow for horizontal scaling; a core advantage of the serverless model. For general
accelerator cards, no comparable feature currently exists. 

Furthermore, it is currently unclear how capable the involved devices in 6G networks will be. Depending on the compute capacity available to the network operator, different functionalities can be more or less suitable to be executed through a machine learning approach. Our proposed solution is therefore not a one-size-fits-all.

The selection of compute nodes can also have a significant impact from a privacy point of view.
While the processing of sensitive data like device location on local devices can greatly reduce
privacy risks, outsourcing this processing to externally managed edge nodes can in turn reintroduce
similar problems already prevalent with external cloud data centers.

\section{Experiments and Results}

In our experimental study, the goal is to demonstrate the benefit and trade-off that the approach of
spreading the workload over a set of 6G nodes has over executing the task on a single cloud node. To
this end, we deploy an ML application in our example architecture on top of two nodes to form a Kubernetes
cluster and connect it to a third node serving as our base station, providing the data. 
We then ensure that two replicas of the function are deployed on two separate nodes to spread the load among them for concurrent requests. We then create two test scenarios to compare:
\begin{enumerate}
    \item Scenario 1: ML/serverless/6G mapping as proposed here, where the nodes are close to each
    other and the ML task is distributed among multiple nodes.
    \item Scenario 2: A centralized cloud-driven approach, where all the ML data is sent to a
    distant cloud node in a data center to execute the ML task in mobile network.
\end{enumerate}

We make use of k-fold cross-validation to determine the best model parameters for each approach and
then use these parameters to train the model on the entire dataset locally.  We then compare the
time it takes to acquire the best parameters as well as the accuracy for the model being trained on
the whole data using the acquired parameters. For fairness, we follow the same serverless paradigm
in both scenarios, differing only regarding the infrastructure and the concurrency; two parallel
requests in the 6G approach (Scenario 1), one request in the mobile network cloud approach (Scenario
2).

\begin{table}[htbp]
\caption{Emulated Network Configuration}
\begin{center}
\begin{tabular}{|c|c|c|c|}
\hline
\textbf{Scenario} & \textbf{Delay}& \textbf{Delay variance}& \textbf{Packet loss} \\
\hline
Scenario 1 & 1.25 ms & 0.25 ms & 0.02\% \\
\hline
Scenario 2 & 15 ms & 3 ms & 0.24\% \\
\hline
\end{tabular}
\label{tab:network}
\end{center}
\end{table}

\begin{figure}[htbp]
    \centering
    \subfloat[Scenario 1: Proposed 6G approach]{\includegraphics[width=0.17\textwidth]{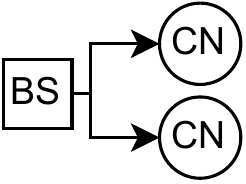}
        \label{fig:exp-edg}}
    \hfil
    \subfloat[Scenario 2: Mobile network cloud approach]{\includegraphics[width=0.26\textwidth]{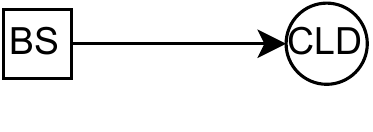}
        \label{fig:exp-cld}}
    \caption{ML Training Approaches}
\end{figure}

In Scenario 1, the dataset will be distributed among two nodes as depicted in
Fig.~\ref{fig:exp-edg}. We therefore split it into three subsets: 20\% of the total set become the
test data, while the remaining data are split into two training sets of 40\% for each instance of
the function respectively. The network emulation capabilities of \texttt{Benchfaster} are used to
configure the network delay, packet loss and jitter of all connections as listed in
Tab.~\ref{tab:network}, thus emulating a wireless 6G environment with low delay and packet loss.

We then utilize a Python script to submit two parallel requests to the framework, resulting in two
training processes. The received output data then contains the accuracy and best parameters of each
training. The better of the two results is then later used to train the model on the full data set, a step that we perform locally for this work, to allow for comparison of the final accuracy of both approaches.

For Scenario 2, also depicted in Fig. \ref{fig:exp-cld}, the data handling procedure is similar.
Here, the dataset is divided into two subsets; 20\% for the test data, while 80\% serves as the
training set, sent as a single request to a cloud node with higher link delay along the connection
as listed in Tab. \ref{tab:network}. The returned parameters are then used to locally train the
model on the training data to acquire a comparable accuracy score.

Therefore, the resulting data that is compared is the final accuracy on the one hand and the time from sending the requests to the serverless function until responses to all requests have been received. Both measurements and tests are repeated 20 times each.

\begin{figure}[htpb]
    \centering
    \subfloat[Response times]{\includegraphics[width=0.23\textwidth]{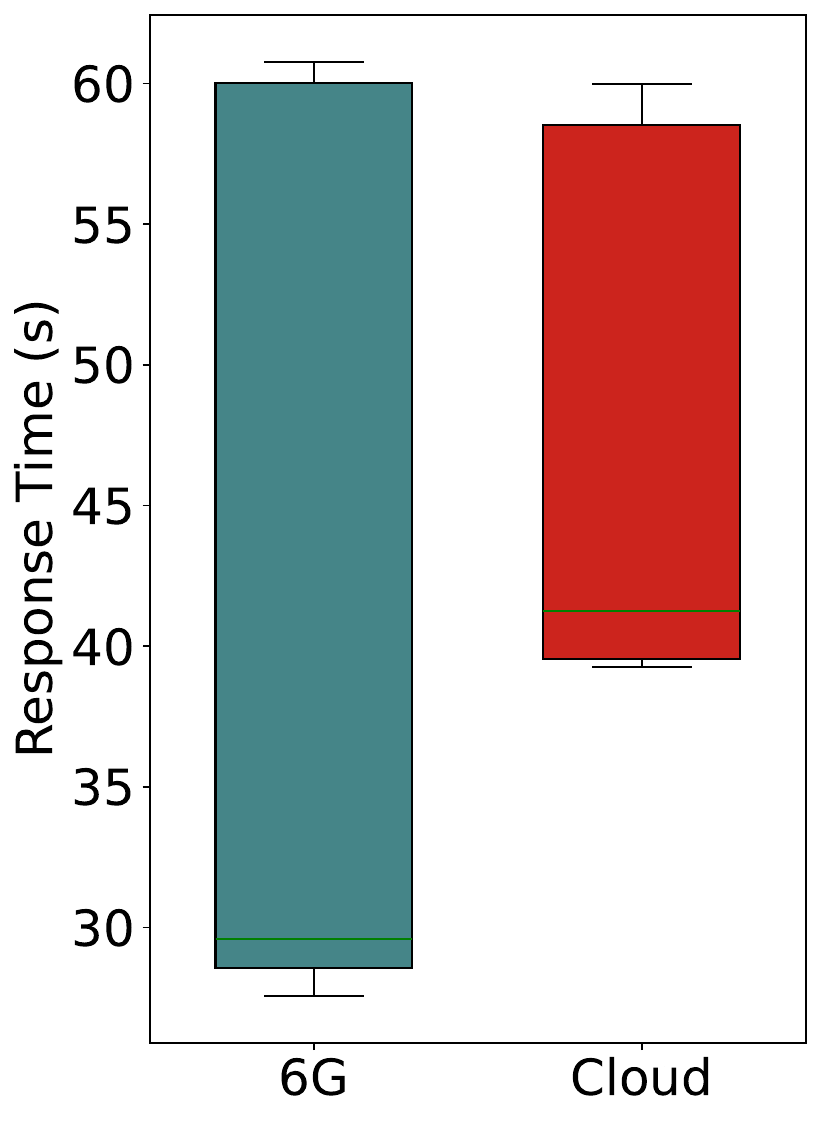}
        \label{fig:result-resp}}
    \hfil
    \subfloat[Test accuracy]{\includegraphics[width=0.23\textwidth]{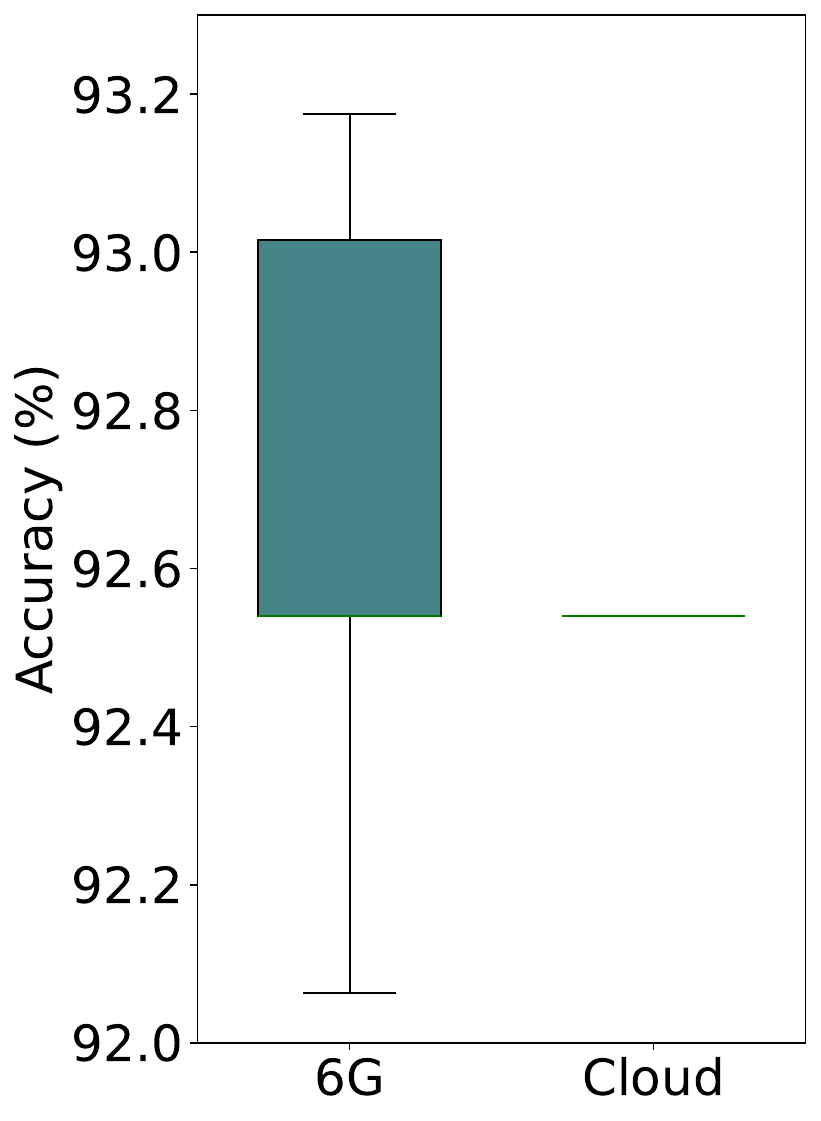}
        \label{fig:result-acc}}
    \caption{Experiment results}
    \label{fig:result}
\end{figure}

The experimental results can be seen in Fig. \ref{fig:result}. It can be observed that the response
times in Fig.~\ref{fig:result-resp} vary to a much greater degree with our proposed 6G approach
(Scenario 1), allowing for mostly significantly lower response times than the cloud can provide
(Scenario 2), but at times also matching or even slightly exceeding comparable response times of the
cloud. Comparing the median values, we can however conclude that in most cases, our proposed
approach improves the response times by around 25\%. It has to be noted that for the task of ML
training, time can have varying criticality depending on the specific 6G applications.

Looking at the test accuracy in Fig.~\ref{fig:result-acc}, we can see a partial improvement for
Scenario 1 over the Scenario 2, with a higher spread around the accuracy delivered by Scenario 2,
while the median values of both approaches remain the same. This is delivered however while only
sending half as much data per request as the cloud did, with a potential to spread this among
different links. Therefore, similar accuracy can be achieved with significantly less link
utilization with the differences varying within a single percent, which makes it insignificant for a
wide variety of use cases. With the dataset sent being identical each time, the Scenario 2 delivers
the exact same accuracy for each request. It has to be noted that the specific accuracy achieved
greatly varies based on the use case; meaning the complexity of data to be classified as well as the
number of training samples provided and the specific machine learning model.

\section{Conclusion}
In this work, we experimentally analyzed the deployment of a class of ML application over serverless
computing in 6G networks. The main motivation was to highlight the possible applications for machine
learning and artificial intelligence in future 6G networks and to show an implementation example of
architecture that utilizes the serverless paradigm. We provided a high-level mapping for ML
applications to serverless to 6G network architectures and analyzed the measurement results in our
proof of concept study. 

Our results show that, while our proposed approach resulted in lower consistency regarding the
response times and accuracy than the traditional mobile network cloud approach, it potentially
allowed for less link utilization if the traffic was split between multiple nodes. This shows that
the serverless approach can be a viable option for upcoming 6G networks, allowing for flexibility
and effective resource utilization as well as load distribution.

Despite the significant implementation effort, the conceptual work is in its infancy, and requires
more in-depth modeling of the mapping process, as well as a richer set of 6G architecture mappings
to drive feasible solutions.



\bibliographystyle{IEEEtran}
\bibliography{refs}

\begin{thebibliography}{10}
\providecommand{\url}[1]{#1}
\csname url@samestyle\endcsname
\providecommand{\newblock}{\relax}
\providecommand{\bibinfo}[2]{#2}
\providecommand{\BIBentrySTDinterwordspacing}{\spaceskip=0pt\relax}
\providecommand{\BIBentryALTinterwordstretchfactor}{4}
\providecommand{\BIBentryALTinterwordspacing}{\spaceskip=\fontdimen2\font plus
\BIBentryALTinterwordstretchfactor\fontdimen3\font minus \fontdimen4\font\relax}
\providecommand{\BIBforeignlanguage}[2]{{%
\expandafter\ifx\csname l@#1\endcsname\relax
\typeout{** WARNING: IEEEtran.bst: No hyphenation pattern has been}%
\typeout{** loaded for the language `#1'. Using the pattern for}%
\typeout{** the default language instead.}%
\else
\language=\csname l@#1\endcsname
\fi
#2}}
\providecommand{\BIBdecl}{\relax}
\BIBdecl

\bibitem{9493388}
V.~P. Rekkas, S.~Sotiroudis, P.~Sarigiannidis, G.~K. Karagiannidis, and S.~K. Goudos, ``Unsupervised machine learning in 6g networks -state-of-the-art and future trends,'' in \emph{2021 10th International Conference on Modern Circuits and Systems Technologies (MOCAST)}, 2021, pp. 1--4.

\bibitem{10396677}
R.~Arshad and R.~Muzzammel, ``Realizing intelligence in 6g communications: A review,'' in \emph{2023 2nd International Conference on Emerging Trends in Electrical, Control, and Telecommunication Engineering (ETECTE)}, 2023, pp. 1--5.

\bibitem{baldoni_data-centric_2023}
\BIBentryALTinterwordspacing
G.~Baldoni, J.~Quevedo, C.~Guimaraes, A.~d.~l. Oliva, and A.~Corsaro, ``Data-centric {Service}-{Based} {Architecture} for {Edge}-{Native} {6G} {Network},'' \emph{IEEE Communications Magazine}, pp. 1--7, 2023, conference Name: IEEE Communications Magazine. [Online]. Available: \url{https://ieeexplore.ieee.org/document/10323415}
\BIBentrySTDinterwordspacing

\bibitem{sabbioni_serverless_2024}
\BIBentryALTinterwordspacing
A.~Sabbioni, A.~Garbugli, L.~Foschini, A.~Corradi, and P.~Bellavista, ``\BIBforeignlanguage{en}{Serverless {Computing} for {QoS}-{Effective} {NFV} in the {Cloud} {Edge}},'' \emph{\BIBforeignlanguage{en}{IEEE Communications Magazine}}, vol.~62, no.~4, pp. 40--46, Apr. 2024. [Online]. Available: \url{https://ieeexplore.ieee.org/document/10364951/}
\BIBentrySTDinterwordspacing

\bibitem{serrano_guest_2024}
\BIBentryALTinterwordspacing
P.~Serrano, C.~Cicconetti, A.~N. Toosi, and M.~Simsek, ``\BIBforeignlanguage{en}{Guest {Editorial}: {Serverless} {Mobile} {Computing}: {From} {Theory} to {Practice}},'' \emph{\BIBforeignlanguage{en}{IEEE Communications Magazine}}, vol.~62, no.~4, pp. 22--22, Apr. 2024. [Online]. Available: \url{https://ieeexplore.ieee.org/document/10494949/}
\BIBentrySTDinterwordspacing

\bibitem{sanchez-artigas_experience_2021}
\BIBentryALTinterwordspacing
M.~Sánchez-Artigas and P.~G. Sarroca, ``\BIBforeignlanguage{en}{Experience {Paper}: {Towards} enhancing cost efficiency in serverless machine learning training},'' in \emph{\BIBforeignlanguage{en}{Proceedings of the 22nd {International} {Middleware} {Conference}}}.\hskip 1em plus 0.5em minus 0.4em\relax Québec city Canada: ACM, Dec. 2021, pp. 210--222. [Online]. Available: \url{https://dl.acm.org/doi/10.1145/3464298.3494884}
\BIBentrySTDinterwordspacing

\bibitem{10209158}
H.~M.~F. Noman, E.~Hanafi, K.~A. Noordin, K.~Dimyati, M.~N. Hindia, A.~Abdrabou, and F.~Qamar, ``Machine learning empowered emerging wireless networks in 6g: Recent advancements, challenges and future trends,'' \emph{IEEE Access}, vol.~11, pp. 83\,017--83\,051, 2023.

\bibitem{muscinelli_overview_2022}
\BIBentryALTinterwordspacing
E.~Muscinelli, S.~S. Shinde, and D.~Tarchi, ``\BIBforeignlanguage{en}{Overview of {Distributed} {Machine} {Learning} {Techniques} for {6G} {Networks}},'' \emph{\BIBforeignlanguage{en}{Algorithms}}, vol.~15, no.~6, p. 210, Jun. 2022, number: 6 Publisher: Multidisciplinary Digital Publishing Institute. [Online]. Available: \url{https://www.mdpi.com/1999-4893/15/6/210}
\BIBentrySTDinterwordspacing

\bibitem{10200390}
T.-H. Nguyen, H.~Park, K.~Seol, S.~So, and L.~Park, ``Applications of deep learning and deep reinforcement learning in 6g networks,'' in \emph{2023 Fourteenth International Conference on Ubiquitous and Future Networks (ICUFN)}, 2023, pp. 427--432.

\bibitem{10489587}
S.~Anjum, D.~Upadhyay, K.~Singh, and P.~Upadhyay, ``Machine learning-based resource allocation algorithms for 6g networks,'' in \emph{2024 2nd International Conference on Disruptive Technologies (ICDT)}, 2024, pp. 1086--1091.

\bibitem{pacheco_towards_2019}
\BIBentryALTinterwordspacing
F.~Pacheco, E.~Exposito, M.~Gineste, C.~Baudoin, and J.~Aguilar, ``Towards the {Deployment} of {Machine} {Learning} {Solutions} in {Network} {Traffic} {Classification}: {A} {Systematic} {Survey},'' \emph{IEEE Communications Surveys \& Tutorials}, vol.~21, no.~2, pp. 1988--2014, 2019, conference Name: IEEE Communications Surveys \& Tutorials. [Online]. Available: \url{https://ieeexplore.ieee.org/document/8543584}
\BIBentrySTDinterwordspacing

\bibitem{10105173}
Y.~Liu, Y.~Deng, A.~Nallanathan, and J.~Yuan, ``Machine learning for 6g enhanced ultra-reliable and low-latency services,'' \emph{IEEE Wireless Communications}, vol.~30, no.~2, pp. 48--54, 2023.

\bibitem{liu2020federated}
Y.~Liu, X.~Yuan, Z.~Xiong, J.~Kang, X.~Wang, and D.~Niyato, ``Federated learning for 6g communications: Challenges, methods, and future directions,'' \emph{China Communications}, vol.~17, no.~9, pp. 105--118, 2020.

\bibitem{10246078}
M.~M. Saeed, R.~A. Saeed, A.~S.~A. Gaid, R.~A. Mokhtar, O.~O. Khalifa, and Z.~E. Ahmed, ``Attacks detection in 6g wireless networks using machine learning,'' in \emph{2023 9th International Conference on Computer and Communication Engineering (ICCCE)}, 2023, pp. 6--11.

\bibitem{10333795}
B.~B. Gupta, K.~T. Chui, A.~Gaurav, and V.~Arya, ``Deep learning based cyber attack detection in 6g wireless networks,'' in \emph{2023 IEEE 98th Vehicular Technology Conference (VTC2023-Fall)}, 2023, pp. 1--5.

\bibitem{wu2020collaborate}
H.~Wu, Z.~Zhang, C.~Guan, K.~Wolter, and M.~Xu, ``Collaborate edge and cloud computing with distributed deep learning for smart city internet of things,'' \emph{IEEE Internet of Things Journal}, vol.~7, no.~9, pp. 8099--8110, 2020.

\bibitem{Carpio2022}
F.~Carpio, M.~Michalke, and A.~Jukan, ``{BenchFaaS}: Benchmarking serverless functions in an edge computing network testbed,'' \emph{{IEEE} Network}, pp. 1--8, 2022.

\bibitem{newscsv}
A.~Kokiantonis, ``News.csv: Data used for a fake news project,'' \url{https://www.kaggle.com/datasets/antonioskokiantonis/newscsv}.

\bibitem{scikit-learn}
F.~Pedregosa, G.~Varoquaux, A.~Gramfort, V.~Michel, B.~Thirion, O.~Grisel, M.~Blondel, P.~Prettenhofer, R.~Weiss, V.~Dubourg, J.~Vanderplas, A.~Passos, D.~Cournapeau, M.~Brucher, M.~Perrot, and E.~Duchesnay, ``Scikit-learn: Machine learning in python,'' \emph{J. Mach. Learn. Res.}, vol.~12, no. null, p. 2825–2830, nov 2011.

\bibitem{bahare_6g_2023}
\BIBentryALTinterwordspacing
M.~K. Bahare, A.~Gavras, M.~Gramaglia, J.~Cosmas, X.~Li, {\"O}.~Bulakci, A.~Rahman, A.~Kostopoulos, A.~Mesodiakaki, D.~Tsolkas, M.~Ericson, M.~Boldi, M.~Uusitalo, M.~Ghoraishi, and P.~Rugeland, ``\BIBforeignlanguage{en}{The {6G} {Architecture} {Landscape} - {European} perspective},'' Feb. 2023, publisher: Zenodo. [Online]. Available: \url{https://zenodo.org/record/7313232}
\BIBentrySTDinterwordspacing

\bibitem{gu_elasticflow_2023}
\BIBentryALTinterwordspacing
D.~Gu, Y.~Zhao, Y.~Zhong, Y.~Xiong, Z.~Han, P.~Cheng, F.~Yang, G.~Huang, X.~Jin, and X.~Liu, ``\BIBforeignlanguage{en}{{ElasticFlow}: {An} {Elastic} {Serverless} {Training} {Platform} for {Distributed} {Deep} {Learning}},'' in \emph{\BIBforeignlanguage{en}{Proceedings of the 28th {ACM} {International} {Conference} on {Architectural} {Support} for {Programming} {Languages} and {Operating} {Systems}, {Volume} 2}}.\hskip 1em plus 0.5em minus 0.4em\relax Vancouver BC Canada: ACM, Jan. 2023, pp. 266--280. [Online]. Available: \url{https://dl.acm.org/doi/10.1145/3575693.3575721}
\BIBentrySTDinterwordspacing

\bibitem{satzke_efficient_2020}
\BIBentryALTinterwordspacing
K.~Satzke, I.~E. Akkus, R.~Chen, I.~Rimac, M.~Stein, A.~Beck, P.~Aditya, M.~Vanga, and V.~Hilt, ``\BIBforeignlanguage{en}{Efficient {GPU} {Sharing} for {Serverless} {Workflows}},'' in \emph{\BIBforeignlanguage{en}{Proceedings of the 1st {Workshop} on {High} {Performance} {Serverless} {Computing}}}.\hskip 1em plus 0.5em minus 0.4em\relax Virtual Event Sweden: ACM, Jun. 2020, pp. 17--24. [Online]. Available: \url{https://dl.acm.org/doi/10.1145/3452413.3464785}
\BIBentrySTDinterwordspacing

\end{thebibliography}

\vfill

\end{document}